\documentclass[10pt,prd,twocolumn,preprintnumbers,showpacs]{revtex4-1}
\usepackage{epsfig}
\usepackage{subfigure}
\usepackage{amssymb}
\usepackage{graphicx}
\usepackage{color}
\usepackage{amsmath}
\usepackage[
       colorlinks=true,
      filecolor=black,
       anchorcolor=blue,
      linkcolor=blue,
      citecolor=blue,
      urlcolor=blue,
       linktocpage=true,
        plainpages=false,
        breaklinks=true,
            pdfstartview=FitH
           ]{hyperref}

\usepackage[dvipsnames]{xcolor}
\usepackage[title]{appendix}
\usepackage{amsmath,epsfig}
\usepackage{amssymb,amsfonts}
\usepackage{latexsym}
\usepackage{bbold}
\usepackage[latin1]{inputenc}
\usepackage{subeqnarray}
\usepackage{xcolor}

\usepackage{graphicx}           
\usepackage{longtable}
\relax
\renewcommand{\theequation}{\arabic{section}.\arabic{equation}}

\begin{document}

\title{No Cauchy Horizon Theorem for Nonlinear Electrodynamics Black Holes with Charged Scalar Hairs} 


\author{Yu-Sen An$^{1,2,3}$}\email{anyusen@itp.ac.cn}
\author{Li Li$^{1,2,4}$}\email{liliphy@itp.ac.cn}
\author{Fu-Guo Yang$^{1,2}$}\email{yangfuguo@itp.ac.cn}

\affiliation{$^1$CAS Key Laboratory of Theoretical Physics, Institute of Theoretical Physics, Chinese Academy of Sciences, P.O. Box 2735, Beijing 100190, China.}

\affiliation{$^2$School of Physical Sciences, University of Chinese Academy of Sciences, No.19A Yuquan Road, Beijing 100049, China.}

\affiliation{$^3$ Department of Physics, Fudan University Shanghai 200433, China}

\affiliation{$^4$School of Fundamental Physics and Mathematical Sciences, Hangzhou Institute for Advanced Study, UCAS, Hangzhou 310024, China.}

\begin{abstract}
We prove a no Cauchy horizon theorem for general nonlinear electrodynamics black holes with charged scalar hairs. By constructing a radially conserved charged, we show that there is no inner Cauchy horizon for both spherical and planar symmetric cases, independent of the form of scalar potential and nonlinear electrodynamics. After imposing the null energy condition, we are also able to rule out the existence of the Cauchy horizon for the hyperbolic black holes. We take the Born-Infeld black hole as a concrete example to study the interior dynamics beyond the event horizon. When the contribution from the scalar potential can be neglected, the asymptotic near-singularity takes a universal Kasner form. We also confirm that the intricate interior dynamics is closely associated with the instability of the inner Cauchy horizon triggered by scalar hairs.
\end{abstract}

\maketitle

\renewcommand{\thefootnote}{\arabic{footnote}}
\setcounter{footnote}{0}


\allowdisplaybreaks


\section{Introduction}
\label{intro}

Black holes have been the object of much interest in both theory and observation. The defining feature of a black hole is the appearance of an event horizon which is the boundary of no escape and divides the spacetime into internal and external regions. While it was well established that black hole formation is a robust prediction of the general theory of relativity~\cite{Penrose:1964wq,Hawking:1969sw}, the convinced observational evidence of black holes has been reported in recent years, including the detection of gravitational waves from merging black holes~\cite{GWs:2016} and the first image of a black hole in the galaxy M87~\cite{shadow:2019I,shadow:2019IV}.  

To understand the interior of a black hole is a fundamental topic in general relativity, which is  important for the studies of black hole physics, gravitation and quantum physics.  As described by general relativity, in the interior of a black hole there typically lies a spacetime singularity that has the property of geodesic incompleteness. In most cases the spacetime curvature becomes infinite at such gravitational singularity inside a black hole. All laws of physics could break down at a singularity and the study of the black hole interior is challenging. Apart from the singularity, there are many black hole solutions which have an inner Cauchy horizon behind the event horizon, \emph{e.g.} Reisser-Nordstr\"{o}m (RN) black hole and Kerr black hole. The Cauchy horizon is defined as the boundary of the domain of dependence of a maximal spacelike hypersurface outside the black hole~\cite{Geroch:1969}, thus is a lightlike boundary of the validity domain of a Cauchy problem. As the region behind the Cauchy horizon can not be predicted by the initial data on the initial maximal spacelike hypersurface, the appearance of Cauchy horizon ruins the predictability of the classical dynamics. The so-called strong cosmic censorship conjecture asks for the instability and ensuing disappearance of Cauchy horizons (see, \emph{e.g.}~\cite{Ringstrom:2015jza,Isenberg:2015rqa}). 

Interestingly, recent development suggests that Cauchy horizons for some kind of charged black holes are nonlinear unstable in the presence of scalar hairs~\cite{Hartnoll:2020rwq,Hartnoll:2020fhc,Cai:2020wrp,Devecioglu:2021xug,Grandi:2021ajl}. In particular, independent of the form of scalar potentials and the asymptotic boundary of spacetimes, a no Cauchy horizon theorem for black holes with charged scalar hairs was proved in~\cite{Cai:2020wrp}. The discussion has been soon extended to black holes in Horndeski theory~\cite{Devecioglu:2021xug} and Gauss-Bonnet theory~\cite{Grandi:2021ajl}. Those studies focus on the charged black holes with the classical Maxwell's electrodynamics. Given its potential importance in understanding the interior structure of black holes, it would be worth classifying other extensions of the theorem. 

In the present work we will extend the theorem to charged black holes with respect to nonlinear generalizations of the Maxwell's electrodynamics and will investigate the interior dynamics of the corresponding hairy black holes.
Nonlinear electrodynamics (NED) has a long history in physics and there are many motivations for the study of NED. For example, the nonlinear Born-Infeld theory~\cite{Born:1934gh} can cure the divergence of self-energy for charged particles and naturally arises in the low energy limit of string theory. The Euler Heinsenberg form~\cite{Heisenberg:1935qt} appears from a 1-loop QED correction to Maxwell's Lagrangian. The unique form which is simultaneously $SO(2)$ electromagnetic duality invariant and conformal has been constructed in~\cite{Bandos:2020jsw}. Interestingly, regular black holes for which the curvature invariants are regular everywhere can be constructed by considering a suitable NED~\cite{AyonBeato:1998ub}, which in the weak field approximation becomes the usual linear Maxwell theory. The NED has also been used to realize the anomalous scalings of strange metals by holographic duality, see \emph{e.g.}~\cite{Kiritsis:2016cpm,Blauvelt:2017koq,Cremonini:2018kla} and attracted many interests in cosmology, see \emph{e.g.}~\cite{Camara:2004ap,Kruglov:2015fbl,Ovgun:2017iwg,Benaoum:2021tec}. There are various theories with nonlinear electromagnetic interactions in the literature and a recent discussion about the black hole thermodynamics in the presence of nonlinear electromagnetic fields can be found in~\cite{Bokulic:2021dtz}.  Note that one can construct a charged black hole with arbitrary number of horizons by considering NED, see \emph{e.g.}~\cite{Nojiri:2017kex,Gao:2017vqv,Hendi:2015esa}. 

Nevertheless, we will show that once a charged scalar hair develops, all inner horizons behind the event horizon are removed for black holes with spherical or planar topological symmetry. For the hyperbolic horizon case, we show that hairy black hole can only have at most one inner horizon with additional (but weak) constraint on the form of NED. Furthermore, if we require the energy momentum tensor to satisfy the null energy condition (NEC), we can prove that there is no Cauchy horizon for hairy black holes with all symmetric horizons.
We will also study the interior dynamics of Born-Infeld action which is natural from a top-down perspective. Similar to the Maxwell theory~\cite{Cai:2020wrp}, the geometry near the singularity takes a Kasner form as long as the contribution of scalar potential to the background is subdominant. In contrast, the gauge potential shows a different scaling behavior from the Maxwell case. Depending on the value of Born-Infeld parameter, the nonlinear dynamics inside the event horizon shows qualitatively different behaviors away from the singularity, depending on whether the normal charged black hole without scalar hair has a Cauchy horizon or not. For the hyperbolic hairy black hole, it is possible to have a fine-tuned Cauchy horizon by violating the NEC. For this case there is a timelike singularity and the asymptotic solution turns out to be a Lifshitz and hyperscaling violating geometry.

The paper is organized as follows. In Section~\ref{Sec:setup}, we introduce the Einstein-NED theory with a charged scalar field and give the equations of motion for isotropic black holes with symmetric horizons, including planar, hyperbolical and spherical topology cases. In Section~\ref{Sec:inner}, we prove the no Cauchy horizon theorem. We study the black hole interior dynamics in the Born-Infeld case for different Born-Infeld parameters in Section~\ref{Sec:BI}. We conclude in Section~\ref{Sec:discussion} with some discussions. In Appendix~\ref{App:Ls} we discuss the constraint on the number of inner horizons for the hyperbolic black holes. In Appendix~\ref{App:h} and Appendix~\ref{App:V}, we consider possible behaviors near the singularity for the hairy Born-Infeld black holes.

\section{Model and Equations of Motion}\label{Sec:setup}
We consider a $(d+2)$ dimensional theory with gravity coupled with an electromagnetic field $A_\mu$ and a charged scalar field $\Psi$. The action reads
\begin{equation}\label{action}
\begin{split}
S&=\frac{1}{2\kappa_N^2}\int d^{d+2}x \sqrt{-g}[\mathcal R+\mathcal L_{M}]\,,\\
\mathcal L_{M}&=L(s)-(D_{\mu}\Psi)^{*}(D^{\mu}\Psi)-V(|\Psi|^2)\,,
\end{split}
\end{equation}
where $L(s)$ is an arbitrary function of the electromagnetic invariant $s=-\frac{1}{4}F_{\mu\nu}F^{\mu\nu}$ with $F_{\mu\nu}=\nabla_\mu A_\nu-\nabla_\nu A_\mu$ and $D_\mu=\nabla_\mu -i q A_\mu$ with $q$ the charge of the scalar field. For the standard Maxwell theory one has $L(s)=s$ which could be thought of as the weak field limit of NED. $V$ is a general function of $|\Psi|^2$ and determines the asymptotic geometry of spacetimes.

The equations of motion derived from the action~\eqref{action} are given by
\begin{equation}\label{EOMs}
\begin{split}
&D_\mu D^\mu \Psi -\dot{V}(|\Psi|^2)\Psi=0 \,,  \\
&\nabla^\mu(L_s F_{\mu\nu} )= i q ( \Psi^* D_\nu \Psi - \Psi D_\nu \Psi^*   ) \,,  \\
& \mathcal{R}_{\mu\nu} -\frac{1}{2}\mathcal{R}g_{\mu\nu}=T_{\mu\nu}\,,
\end{split}
\end{equation}
with the energy momentum tensor
\begin{equation}
\begin{split}
T_{\mu\nu}=\frac{1}{2}\left[ D_\mu \Psi (D_\nu \Psi)^* + D_\nu \Psi (D_\mu \Psi)^* \right]\\
+ \frac{L_s}{2}F_{\mu\sigma}{F_\nu}^\sigma+\frac{1}{2}\mathcal{L}_M g_{\mu\nu}\,,
\end{split}
\end{equation}
where $\dot{V}(x)=d V(x)/dx$ and $L_s=d L(s)/ds$.

We consider the static hairy black holes with symmetric horizons, for which the generic ansatz is given by~\cite{Cai:2020wrp}
\begin{equation}\label{ansatz}
\begin{split}
ds^2=\frac{1}{z^2}\left[-f(z)e^{-\chi(z)}dt^2+\frac{dz^2}{f(z)}+d\Sigma^2_{d,k}\right]\,,\\
A_{\mu}dx^{\mu}=A_{t}(z)dt, \quad \Psi=\psi(z)\,,
\end{split}
\end{equation}
where $d\Sigma^2_{d,k}$ denotes the line element of $d$-dimensional unit sphere $(k=1)$, flat plane $(k=0)$ or unit hyperbolic plane $(k=-1)$.

Substituting the ansatz~\eqref{ansatz} into the equations of motion~\eqref{EOMs}, we obtain the following independent equations 
\begin{equation}\label{eom-at}
z^d(\frac{e^{\chi/2}}{z^{d-2}}L_{s}A_{t}')'=\frac{2e^{\chi/2}}{f}q^2\psi^2A_{t}\,,
\end{equation}
\begin{equation}\label{eom-psi}
\begin{aligned}
z^{d+2}e^{\chi/2}(\frac{e^{-\chi/2}}{z^{d}}f\psi')'=\dot V\psi-\frac{z^2e^{\chi}}{f}q^2A_{t}^2\psi\,,
\end{aligned}
\end{equation}
\begin{equation}\label{eomchi}
\frac{d}{2}\chi' =z\psi'^2+\frac{z e^{\chi} }{f^2}q^2 \psi^2A_t^2\,,
\end{equation}
\begin{equation}\label{eomf}
\begin{aligned}
dz^{d+2}(z^{-d-1}f)'=V-&L(s)-kd(d-1)z^2\\
&+\frac{d}{2}zf\chi'+L_{s}z^4e^{\chi}A_t'^2.
\end{aligned}
\end{equation}
Here the prime denotes the derivative with respect to the radial coordinate $z$.

\section{Interior structure}\label{Sec:inner}
In this section we will understand the interior structure of the hairy black holes~\eqref{ansatz} for the theory~\eqref{action} with NED. We find two methods to remove the inner Cauchy horizon. The first method is based on the conserved charge~\cite{Cai:2020wrp} and the second one takes advantage of the NEC.

\begin{figure}[h!]
\includegraphics[width=0.45\textwidth]{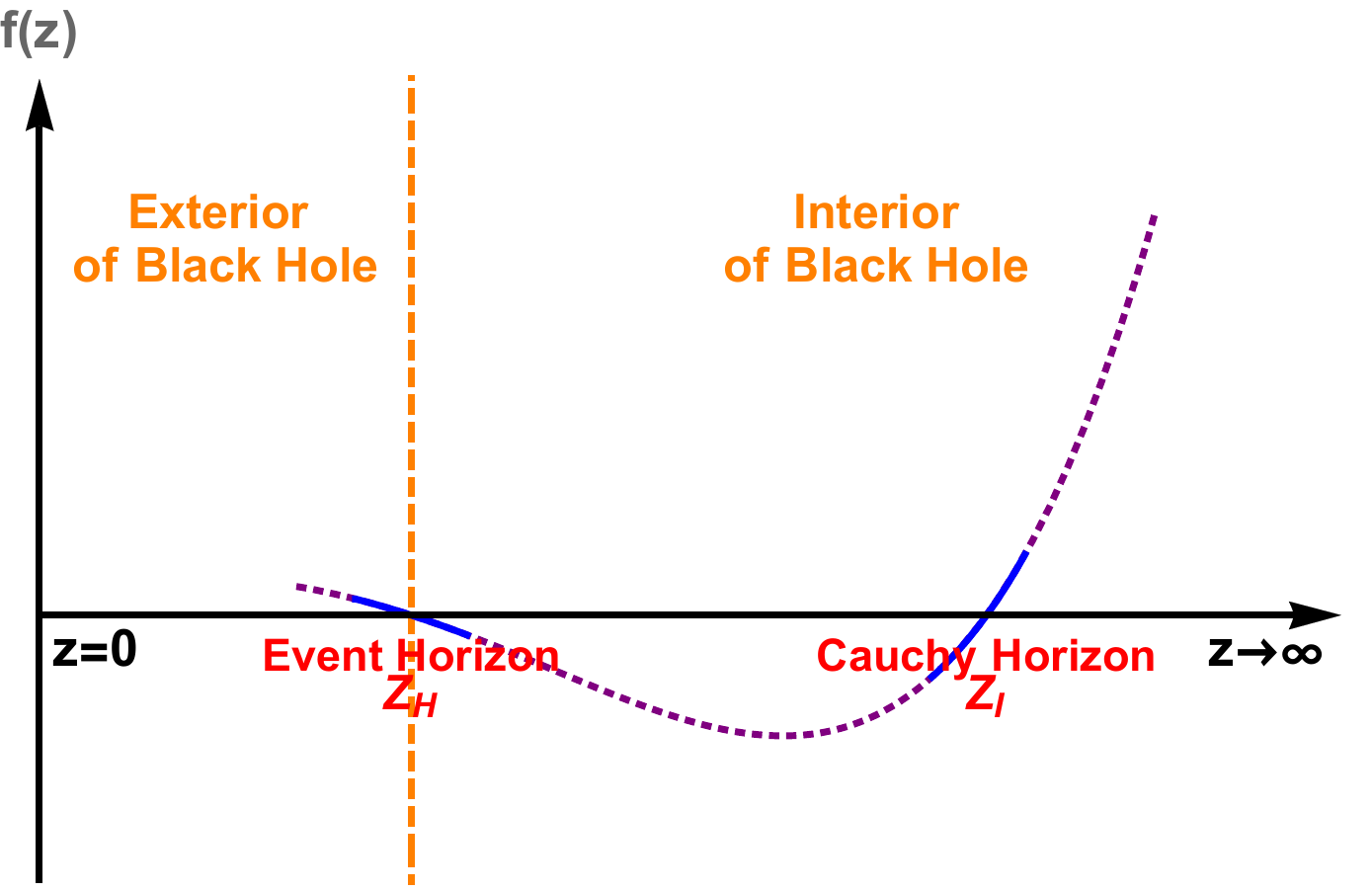}
  \caption{Schematic structure for the blackening function $f(z)$ of a black hole at finite temperature with an event horizon at $z_H$ and an inner Cauchy horizon at $z_I$. $f(z)$ is vanishing at both horizons with $f'(z_H)<0$ and $f'(z_I)> 0$ in the coordinate system we are considering.}
  \label{fig:fig0}
\end{figure}
Besides the event horizon at $z_H$, we assume that there is an inner Cauchy horizon at $z_I$ behind the event horizon. Then one has $f(z_H)=f(z_I)=0$ with $z_I>z_H$. We consider a black hole at finite temperatures, for which the blackening function $f(z)$ turns from positive to negative near $z_H$ towards the interior. Near the inner horizon $z_I$, $f(z)$ turns from negative to positive for which $f'(z_I)>0$~\cite{footnote0}. The schematic structure for this case with two horizons is illustrated  in Fig.~\ref{fig:fig0}. Therefore, we have
\begin{equation}\label{cond1}
\begin{split}
f(z_H)=&0,\quad f'(z_H)<0\,,\\
f(z_I)=&0,\quad f'(z_I)>0\,.
\end{split}
\end{equation}
Furthermore, for a black hole with non-trivial charged scalar hairs, the smoothness of geometry near both horizons yields~\cite{Cai:2020wrp} 
\begin{equation}\label{cond2}
A_t(z_H)=0,\quad A_t(z_I)=0\,.
\end{equation}

\subsection{Proof of no Cauchy horizon via conserved charge}\label{sec:Q}
Following~\cite{Cai:2020wrp}, we first prove the no Cauchy horizon theorem for planar and spherical topological horizons without knowing the details of the interior geometry. Then we consider the hyperbolic case, for which we can constraint the number of inner horizons under particular condition.

Our first task is to extend the radially conserved charge $\mathcal{Q}$ of~\cite{Cai:2020wrp} to the NED case. Using the equations of motion, we obtain the following radially conserved quantity
\begin{equation}\label{Q}
\begin{aligned}
\mathcal{Q}(z)=&z^{2-d}e^{\chi/2}[z^{-2}(fe^{-\chi})'-L_{s}A_{t}A_{t}']\\
	     &+2k(d-1)\int^z y^{-d}e^{-\chi(y)/2}dy\,,
\end{aligned}
\end{equation}
for which $\mathcal{Q}'(z)=0$. Note that the last integrated term depends on the topology of a black hole.

Taking advantage of $\mathcal{Q}(z_H)=\mathcal{Q}(z_I)$ and using the conditions~\eqref{cond1} and~\eqref{cond2}, we obtain
\begin{equation}\label{qhi}
\begin{aligned}
\frac{f'(z_{H})}{z_{H}^{d}}&e^{-\chi(z_{H})/2}-\frac{f'(z_{I})}{z_{I}^{d}}e^{-\chi(z_{I})/2}\\
&\qquad\;\;=2k(d-1) \int_{z_{H}}^{z_{I}} y^{-d}e^{-\chi(y)/2} dy,
\end{aligned}
\end{equation}
with $z_H$ and $z_I$ representing the location of event horizon and Cauchy horizon, respectively. It is obvious that the left hand side of~\eqref{qhi} is negative since $f'(z_{H})<0$ and $f'(z_{I})>0$ (see Fig.~\ref{fig:fig0}). On the other hand, the integration of the right hand side is positive as its integrand is positive for any $z$. As a result, we have the following three cases.

\subsubsection{Spherical topology case \texorpdfstring{$k=1$}{TEXT}}
For the spherical topology, the right hand side of~\eqref{qhi} is positive, while the left hand side is negative. Therefore, smooth Cauchy horizon is forbidden for spherical black holes with charged scalar hairs in the presence of NED. 

\subsubsection{Planar topology case \texorpdfstring{$k=0$}{TEXT}}
For the planar black hole ($k=0$), the right hand side of~\eqref{qhi} is vanishing, while the left hand side is negative. Again, there is no smooth Cauchy horizon for planar black holes with charged scalar hairs. 

\subsubsection{Hyperbolic topology case \texorpdfstring{$k=-1$}{TEXT}}
For the hyperbolic case with $k=-1$, both sides of~\eqref{qhi} are negative. Therefore, we are not able to rule out the existence of an inner Cauchy horizon. In contrast to both spherical and planar cases, it is possible to develop an inner horizon as long as~\eqref{qhi} is satisfied. In fact, a concrete example for the hyperbolic black hole with an inner horizon was provided in~\cite{Cai:2020wrp} for $L(s)=s$. For the hyperbolic case with the standard Maxwell term, it have been proved that the hairy black holes can only have at most one inner horizon~\cite{Cai:2020wrp}. However, the argument of~\cite{Cai:2020wrp} does not apply to the present case with general NED \emph{i.e.} $L(s)$. Only for the NED with $L_s\geq0$ can one use the method developed by~\cite{Cai:2020wrp} and find that there is at most one inner horizon, see Appendix~\ref{App:Ls} for more details. 

\subsection{NEC removes Cauchy horizons}\label{sec:NEC}

Note, however, that in above argument we do not impose any constraint on the form of matter fields. More recently, the authors of~\cite{Yang:2021civ} found that if the interior of a static black hole with hyperbolic or planar symmetry is dominated by classical matter satisfying NEC, there is at most one non-degenerate inner horizon behind the event horizon. Interestingly, we now show that for the present theory~\eqref{action} we can give a much stronger constraint on the number of inner horizons by imposing the NEC. The NEC says that $T_{\mu\nu}k^\mu k^\nu\geq0$ for arbitrary null vector $k^\mu$. For general $L(s)$, we can prove that there is no inner Cauchy horizon provided the NEC is satisfied.  

The proof is given as follows. Since there is an inner horizon at $z_I$, the blackening function $f(z)<0$ between $z_H$ and $z_I$ (Fig.~\ref{fig:fig0}). For our present theory with the ansatz~\eqref{ansatz}, the NEC implies
\begin{equation}\label{null}
T_{ab}k_0^a k_0^b=\frac{1}{2}z^d e^{\chi/2}\left[\frac{2e^{\chi/2} q^{2}\psi^{2}A_{t}^{2}}{f z^{d}}+\frac{e^{\chi/2}}{z^{d-2}} L_{s} A_{t}'^{2}\right]\geq0\,,
\end{equation}
where we have chosen the null vector $k_0^a=\sqrt{-f}(\frac{\partial}{\partial z})^a+\frac{1}{z \sqrt{g_{ii}}}(\frac{\partial}{\partial x^i})^a$ with $x^i$ one of the coordinates of the spacelike cross-section of black hole event horizon.

Then we obtain that 
\begin{eqnarray}\label{nullAt}
\left(\frac{e^{\chi/2}}{z^{d-2}}L_{s} A_{t}' A_{t}\right)'&=&\frac{2e^{\chi/2} q^{2}\psi^{2}A_{t}^{2}}{f z^{d}}+\frac{e^{\chi/2}}{z^{d-2}} L_{s} A_{t}'^{2}\,,\nonumber\\
&=&\frac{2e^{-\chi/2}}{z^d}T_{\mu\nu}k_0^\mu k_0 ^\nu\geq0\,,
\end{eqnarray}
where in the first equality we have used~\eqref{eom-at}. Therefore, the combination $\frac{e^{\chi/2}}{z^{d-2}}L_{s} A_{t}' A_{t}$ is a monotonic increasing function of $z$. Since $A_t(z_H)=0$, we should have 
$\frac{e^{\chi/2}}{z^{d-2}}L_{s} A_{t}' A_{t}>0$ inside the event horizon $z>z_H$ for non-trivial $L_s$. If there is an inner horizon at $z_I>z_H$, we will have $f(z_I)=0$ and, more importantly, $A_t(z_I)=0$ due to the smoothness of geometry. This is not possible as $\frac{e^{\chi/2}}{z^{d-2}}L_{s} A_{t}' A_{t}>0$ at $z_I$. Thus, the NEC necessary removes the Cauchy horizon of black holes with charged scalar hairs.

\section{Dynamics inside Born-Infeld Black Hole}\label{Sec:BI}
So far we have completed the proof of the theorem for the interior structure of hairy black holes with NED. We have shown that the inner horizon is in general absent and the black hole interior ends at a spacelike singularity or endpoint as $z\rightarrow\infty$. In this section, we are interested in the interior dynamics of the hairy black holes. It has been shown that the behavior will be sensitive to the details of the model because of the dynamics in highly nolinear regime~\cite{Hartnoll:2020rwq,Hartnoll:2020fhc,Cai:2020wrp,Devecioglu:2021xug,Grandi:2021ajl}. Among all theories with NED, Born-Infeld case is the most well known nonlinear model and is quite natural from string theory. Therefore, in the present paper we will focus on the Born-Infeld theory.

The action reads
\begin{equation}\label{BI}
L(s)=\beta^{2}\left(1-\sqrt{1-\frac{2s}{\beta^{2}}}\right),
\end{equation}
with $\beta$ the Born-Infeld parameter. As $\beta \to \infty$, we obtain the linear Maxwell term $L=s=-\frac{1}{4}F_{\mu\nu}F^{\mu\nu}$. Note also that
\begin{equation}
L_s=\frac{1}{\sqrt{1-\frac{2s}{\beta^2}}}\geqslant 0\,.
\end{equation}
For future convenience, we introduce the following quantity
\begin{equation}\label{echarge}
\rho=L_{s} z^{2-d}e^{\chi/2} A_{t}'\,,
\end{equation}
which characterizes the charge degrees of freedom behind the surface generating a nonzero electric flux in the deep interior.

The equations of motion now become
\begin{equation}\label{eqh}
h'=\frac{e^{-\chi/2}}{2}\left(-\frac{2k}{z^{2}}+\frac{\rho^{2}}{L_s}+\frac{V(\psi^2)-L(s)}{z^{4}}\right)\,,
\end{equation}
\begin{equation}\label{eqchi}
\chi'=z\psi'^{2}+\frac{\psi^{2}A_{t}^{2}q^{2}}{h^{2}z^{5}}\,,
\end{equation}
\begin{equation}\label{eqpsi}
\psi''=-\left(\frac{h'}{h}+\frac{1}{z}\right)\psi'+\left(\frac{\dot V e^{-\chi/2}}{z^5h}-\frac{A_{t}^{2}q^{2}}{h^{2}z^{6}}\right)\psi\,,
\end{equation}
\begin{equation}\label{eqrho}
\rho'=\frac{2q^{2}\psi^{2}A_{t}}{z^{5}h}\,,
\end{equation}
where we have introduced a new function $h=f e^{-\chi/2}z^{-3}$ and have set $d=2$ (the generalization to other dimensions is straight forward).

\subsection{Kasner singularity}\label{sub:kasner}
In order to know explicitly the interior geometry, one needs to solve the coupled equations of motion from~\eqref{eqh} to~\eqref{eqrho} numerically.
The divergency near the singularity $z\rightarrow \infty$ challenges the numerics. So we first adopt an analytic approach about the asymptotic near-singularity behavior.

We will consider the case for which the scalar potential $V$ can be neglected. Since there is no Cauchy horizon, we have $h(z)<0$ inside the event horizon and
\begin{equation}\label{hh}
\lim_{z\rightarrow\infty}h(z)=-h_0<0\,,
\end{equation}
with $h_0$ a finite positive constant. Hence, the $h'/h$ term of the scalar equation~\eqref{eqpsi} can be neglected at large $z$. The other possibility for which $h(z)$ diverges near the singularity can be excluded, see Appendix~\ref{App:h} for more details. Thanks to the structure of Born-Infeld action~\eqref{BI} for which $s<\beta^2/2$, we obtain that
\begin{equation}\label{maxAt}
A_{t}'^{2} < \frac{e^{-\chi} \beta^{2}}{z^{4}}\,.
\end{equation}
Since $\chi$ is a monotonic increasing function, see~\eqref{eqchi}, and is finite at $z_H$, we find that $e^{-\chi/2}\leqslant \mathcal{O}(z^0)$. Therefore, we have
\begin{equation}
|A_{t}'|<{\beta e^{-\chi/2}}/{z^{2}}\leqslant \mathcal{O}(1/z^2)\,,
\end{equation}
and  $|A_{t}|$ is at most of $\mathcal{O}(z^0)$.  

Thus, the last term of the scalar equation~\eqref{eqpsi} can be neglected near the singularity and one has 
\begin{equation}
\psi''=-\frac{1}{z}\psi'\,.
\end{equation}
Therefore, at large $z$ the scalar field $\psi$ is given by
\begin{equation}
\psi(z)=\sqrt{2}\alpha \ln(z),
\end{equation}
with $\alpha$ a constant. Substituting $\psi(z)$ into~\eqref{eqchi} and noting that $A_t$ is at most a constant, we find that $\chi'=2\alpha^2/z$ which yields
\begin{equation}
\chi(z)=2\alpha^2 \ln(z).
\end{equation}
By using $h=fe^{-\chi/2} z^{-3}\sim -h_0$, we then obtain 
\begin{equation}
f(z) = -f_s z^{3+\alpha^{2}}+\cdots\,,
\end{equation}
with $f_s$ a positive constant.

Let's turn to the gauge potential. When $z$ goes to infinity, we have
\begin{equation}\label{q'lim}
|\rho'|=|\frac{2q^{2}\psi^{2}A_{t}}{z^{5}h}| \leqslant \mathcal{O} \left(\frac{\ln (z)^2}{z^5}\right)\rightarrow 0\,.
\end{equation}
As a result, at large $z$, we have
\begin{equation}\label{rho}
\rho=L_{s} e^{\chi/2}A_{t}'=\frac{e^{\chi/2}A_t'}{\sqrt{1-\frac{z^4 e^\chi A_t'^2}{\beta^2}}}\approx \rho_0,
\end{equation}
with $\rho_0$ a constant. Solving~\eqref{rho}, we obtain 
\begin{equation}
A_{t}'= \beta z^{-2-\alpha^{2}}+\dots\,.
\end{equation}
For comparison, $A_t'\sim z^{-\alpha^2}$ for the linear Maxwell case~\cite{Cai:2020wrp}, corresponding to $\beta\rightarrow\infty$. One can see that $A_t'$ of hairy Born-Infield black holes decays much faster than the Maxwell theory. This is due to the fact that the Born-Infeld action~\eqref{BI} contains a square root which constraints strongly the maximum value allowed for $A_t'$, see~\eqref{maxAt}. In contrast, there is, a priori, no strong constraint on $A_t'$ in the Maxwell case.

To sum up, the asymptotic geometry near the spacelike singularity takes the following form:
\begin{equation}\label{kasner}
\begin{split}
&ds^{2}=\frac{1}{z^{2}}\left[f_s z^{3-\alpha^{2}}dt^{2}-\frac{1}{f_s z^{3+\alpha^{2}}}dz^{2}+d\Sigma^{2}_{2,k}\right]\,,\\
&\psi(z)=\sqrt{2}\alpha \ln(z),\quad A_{t}'=\beta z^{-2-\alpha^{2}}\,.
\end{split}
\end{equation}
One finds that all metric components are power laws and the scalar field is logarithmic.
By the coordinate transformation $\tau \sim z^{-(3+\alpha^{2})/2}$, we obtain 
\begin{equation}
\begin{split}
&ds^{2}=-d\tau^{2}+c_{t}\tau^{2p_{t}} dt^{2}+c_{s} \tau^{2p_{s}} d\Sigma^{2}_{2,k}\,,\\
&\psi(\tau)= -p_\psi\ln(\tau)\,,
\end{split}
\end{equation}
with
\begin{equation}
p_{t}=\frac{\alpha^{2}-1}{\alpha^{2}+3},\ p_{s}=\frac{2}{\alpha^{2}+3},\ p_{\psi}=\frac{2\sqrt{2}\alpha}{\alpha^{2}+3}\,,
\end{equation}
The geometry takes the Kasner form with the following relation
\begin{equation}
p_{t}+2p_{s}=1,\ p_{t}^{2}+2 p_{s}^{2}+p_{\psi}^{2}=1\,.
\end{equation}

Hence, while the scaling of $\chi$, $\psi$ and $f$ is the same as the Maxwell case of~\cite{Cai:2020wrp}, the scaling of $A'_{t}$ has a significant change due to the nonlinear modification of the electromagnetic sector. Note that to obtain the Kasner behavior we have neglected the contribution from the scalar potential.  Therefore, one should have the following constraint at large $z$:
\begin{equation}\label{kconstraint}
\frac{|V(\psi^2)|}{z^{3+\alpha^2}}\ll 1\,,
\end{equation}
which allows the potential $V$ to be a general algebraic function, including polynomial functions (see also Appendix~\ref{App:V}). As shown in~\cite{Cai:2020wrp}, when the scalar potential becomes important to the background (\emph{e.g.} exponential form of potential), novel behaviors different from the Kasner form could appear. We will leave this for future study. For the hyperbolic topology, the hairy black hole with a Cauchy horizon is possible by fine tuning parameters, for which the resulted singularity is timelike. We will return to this particular case in Section~\ref{Sec:discussion}.

\subsection{Interior dynamics before Kasner singularity}
Apart from the Kasner singularity, we now consider other intricate dynamical behaviors before reaching the singularity. There are several intermediate regimes for which we will study numerically. We are interested in the effect of the Born-Infeld parameter $\beta$ on the interior dynamics. We consider some explicit examples with the spherical topology ($k=1$) and the scalar potential $V=-6+\Psi^{2}$. It allows the charged black holes with asymptotically anti-de Sitter (AdS) boundary. We also fix $q=z_H=1$.

For the case without scalar hairs, we can solve the system analytically and the resulted Born-Infeld AdS black hole is given by
\begin{equation}\label{BIbh}
\begin{split}
f(z) = 1&+z^2+M z^3+\frac{\beta^2}{6}\\
&-\frac{\beta^2}{6}{_2}F{_1} \left( -\frac{3}{4},-\frac{1}{2};\frac{1}{4};-\frac{\rho^2 z^4}{\beta^2}\right)\,,\\
 A_t(z) = \mu&-\rho\, z \;{_2}F_{1}\left(\frac{1}{4},\frac{1}{2};\frac{5}{4};-\frac{z^4}{\beta^2}\right)\,,
 \end{split}
\end{equation}
with $\chi=\psi=0$. Here the constant $M$ is related to the mass of the black hole. $\rho$ has been given in~\eqref{echarge} and is a constant for the hairless case ($\chi=\psi=0$) for $d=2$, accounting for the charge density of the background. $\mu$ is the chemical potential which can be fixed in such a way that $A_t$ is vanishing at the event horizon $z_H$. We call~\eqref{BIbh} the normal solution since it dose not break the local U(1) gauge symmetry of~\eqref{action}. For the other case with scalar hairs, one has to solve the coupled equations of motion~\eqref{eqh}-\eqref{eqrho} numerically.

\begin{figure}[h!]
\includegraphics[width=0.42\textwidth]{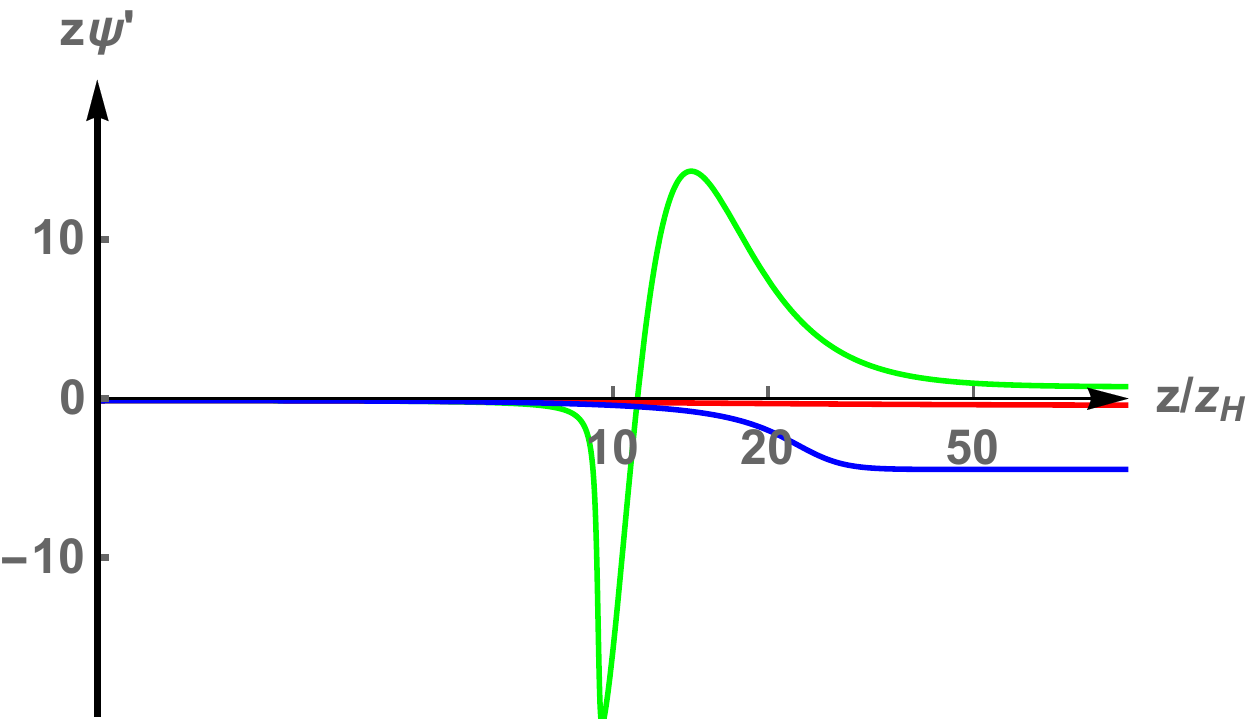}
\includegraphics[width=0.40\textwidth]{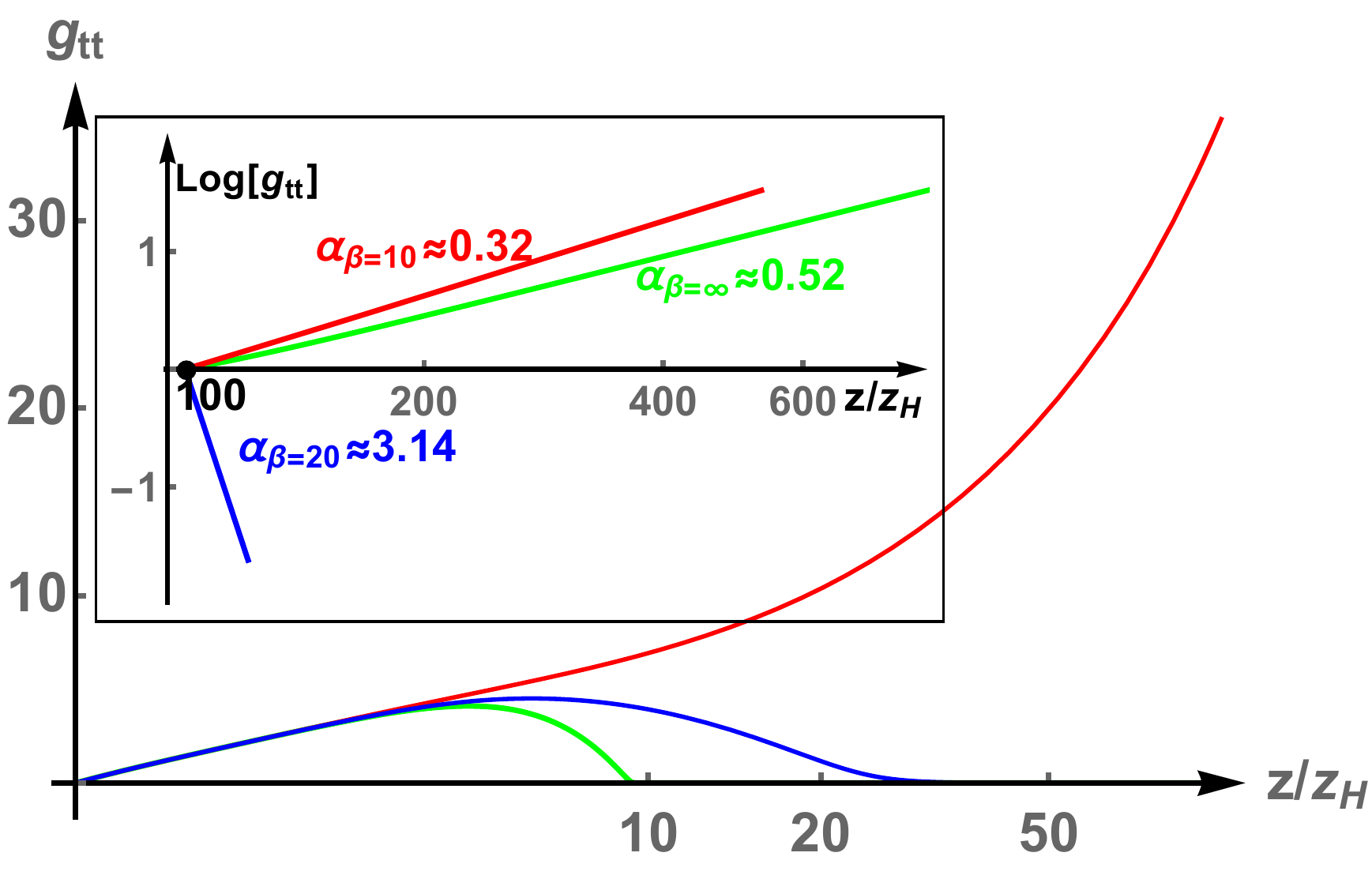}\\
\includegraphics[width=0.40\textwidth]{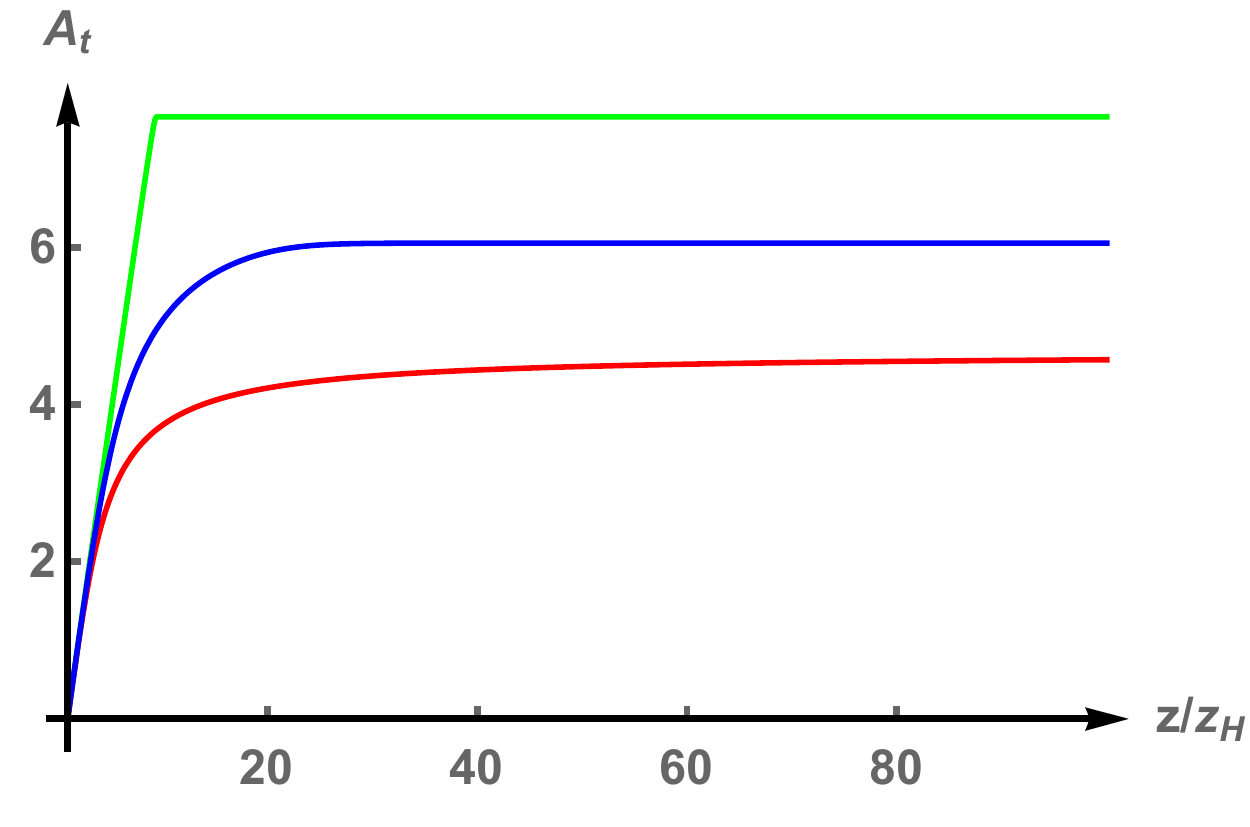}\\
  \caption{Comparison of $z\psi'$ (up), $g_{tt}=-f e^{-\chi}/z^2$ (middle) and $A_t$ (bottom) for different values of Born-Infeld parameter $\beta$. The green curve corresponds to the linear Maxwell case $\beta \to \infty$, while the blue and red curves correspond to the Born-Infeld balck hole with $\beta=20$ and $\beta=10$, respectively. See the insert for the behaviors of $g_{tt}$ at large $z$, where we have moved all curves to the same starting point at $z/z_H=100$ for comparing the Kasner behavior. Numerics are with $A_t(z_H)=\chi(z_{H})=0$, $A_{t}'(z_{H})=1$, $\psi(z_{H})=0.5$ and $q=1$. Other boundary conditions at $z_H$ can be fixed by the series expansion near the event horizon. }
  \label{Fig:field}
\end{figure}

First, we integrate the system from the event horizon $z_H$ to the deep interior by choosing a large value of $\psi$ at $z_H$. A large value of $\psi(z_H)$ is to suppress the potential intricate dynamics associated with the strong nonlinear behaviors near the ``would-be" inner horizon of the Born-Infeld balck hole~\eqref{BIbh}~\cite{Hartnoll:2020fhc,Cai:2020wrp}. A comparison of interior behaviors for different values of Born-Infeld parameter $\beta$ is presented in Fig.~\ref{Fig:field}. Apart from confirming the Kasner form~\eqref{kasner} near the singularity, one finds that the dynamics inside the event horizon depends on $\beta$.

In particular, the exponent $\alpha$ of the Kasner singularity~\eqref{kasner} is sensitive to $\beta$, which can be seen from  the first plot of Fig.~\ref{Fig:field} (note that $z\psi'= \sqrt{2}\alpha$ at large $z$ limit). For $g_{tt}$, one can see that it becomes positive inside the black hole and first increases as $z$ is increased from $z_H$. Depending on the value of $\beta$, $g_{tt}$ will continue increasing or turn to be decreasing when $z$ is sufficiently large. Note that $g_{tt}\sim z^{1-\alpha^2}$ in the large $z$ regime. For the blue curve $\alpha>1$ and therefore $g_{tt}$ finally decreases rapidly. For the green and red curves, $\alpha<1$ and $g_{tt}$ blows up, see the middle plot. In the last plot of Fig.~\ref{Fig:field}, we compare the gauge potential for different $\beta$. One can find that $A_{t}$ saturates to a constant very quickly. This can be understood as follows. As $\chi(z)$ becomes very large with respect to $z$, to avoid making the Born-Infeld action imaginary, $A_{t}'(z)$ should quickly flow to zero. For the present case, $A_t$ saturates to its maximum value more quickly as the parameter $\beta$ is increased, and  the maximum value is larger for larger $\beta$.

Next, we study the possible crossover that occurs near the ``would-be" inner horizon of~\eqref{BIbh}. More precisely, the highly nonlinear dynamics in this regime would lead to the so-called collapse of Einstein-Rosen bridge for which $g_{tt}$ shrinks very rapidly and Josephson oscillations for which the scalar field undergoes rapid oscillations. Such intricate dynamics was argued to be due to the instability of the would-be inner Cauchy horizon triggered by nonzero scalar field~\cite{Hartnoll:2020rwq,Hartnoll:2020fhc}. The instability was found to be stronger for small values of the scalar, uncovering the nonlinear nature of the dynamics.
In contrast to the Maxwell case, the normal Born-Infeld black hole~\eqref{BIbh} can be RN like with an inner Cauchy horizon and a timelike singularity or Schwarzschild like for which there is no inner horizon and the singularity is spacelike, depending on the Born-Infeld parameter $\beta$. So it is a perfect playground for us to investigate the relation between the instability of inner horizon of~\eqref{BIbh} and the intricate dynamics just mentioned above. 

\begin{figure}[h!]
\includegraphics[width=0.45\textwidth]{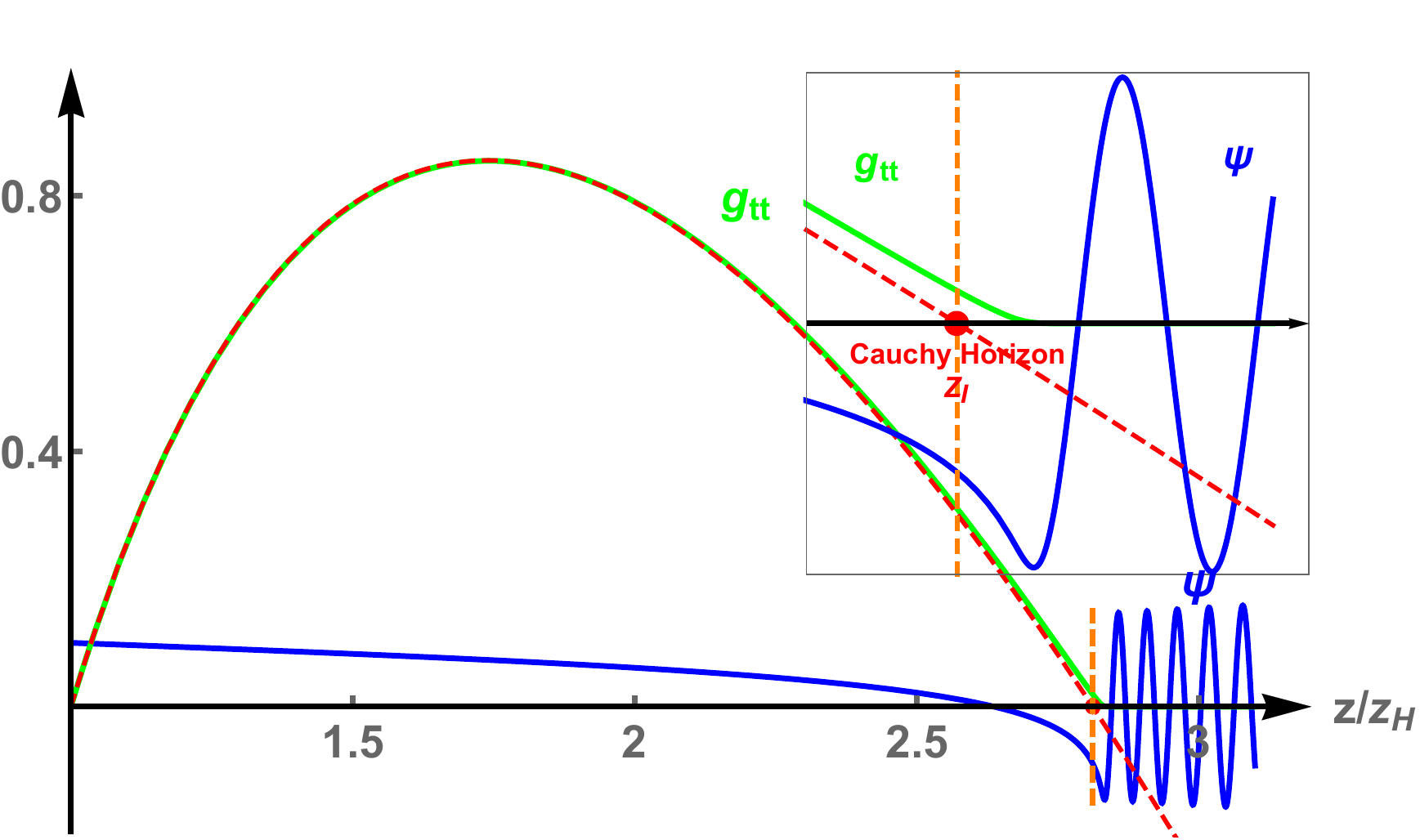}
 \caption{Interior dynamics for the hairy Born-Infeld black hole for $\beta=8$ and $A_{t}'(z_{H})=2$. The red dashed line is the function $g_{tt}$ of the normal Born-Infeld black hole which has a Cauchy horizon at $z_I\approx2.812z_H$. The green and blue curves denote $g_{tt}$ and $\psi$ of the hairy black hole, respectively. The collapse of Einstein-Rosen bridge and Josephson oscillations of the scalar field are obvious near the Cauchy horizon $z_I\approx2.812z_H$ of the normal solution.}
 \label{Fig:Cauchy}
\end{figure}
\begin{figure}[h!]
\includegraphics[width=0.43\textwidth]{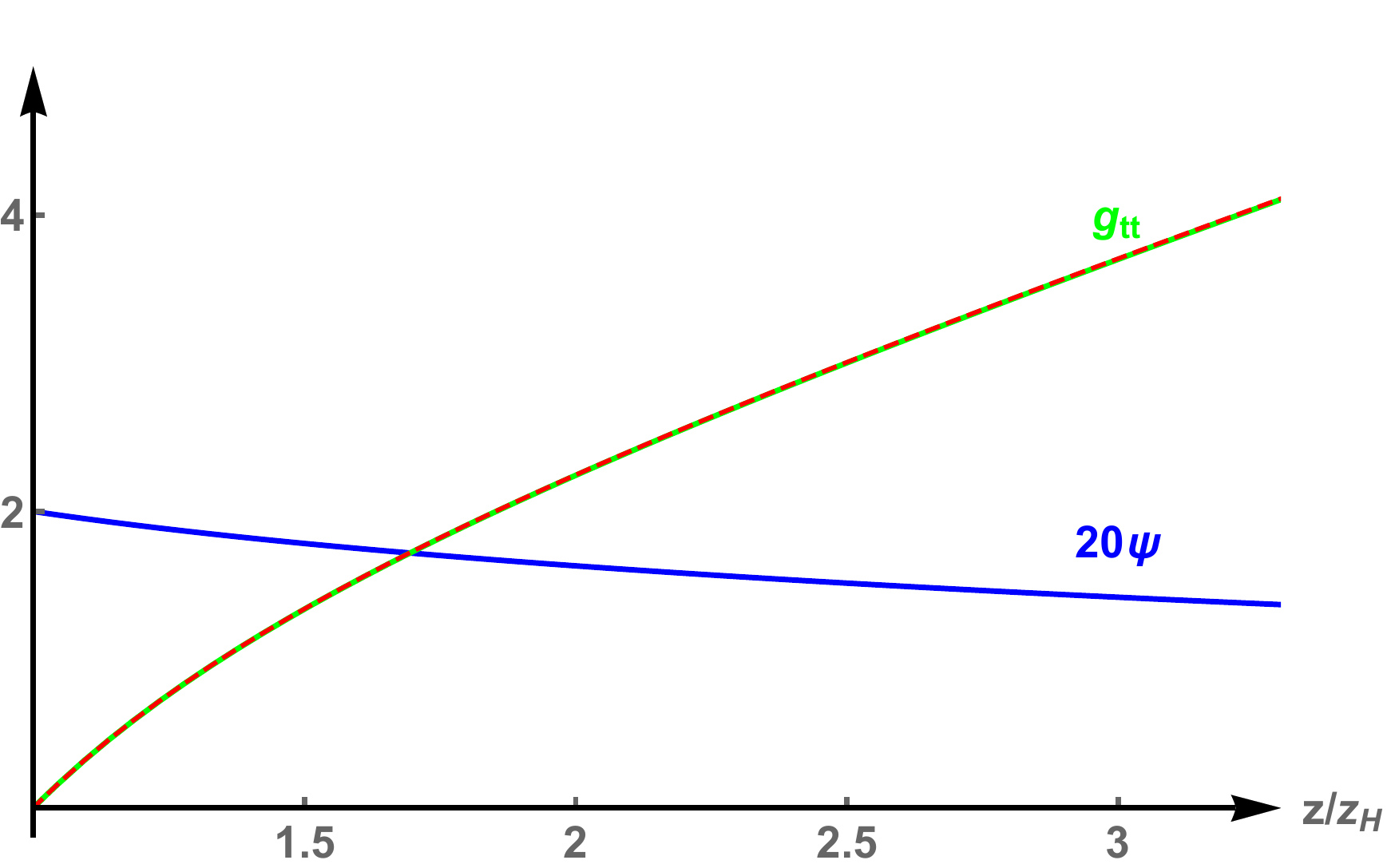}\\
 \caption{Interior dynamics for the hairy Born-Infeld black hole for $\beta=2$ and $A_{t}'(z_{H})=1$. The red dashed line denotes $g_{tt}$ of the hairless Born-Infeld black hole without Cauchy horizon. The green and blue curves correspond to $g_{tt}$ and $\psi$ of the hairy black hole, respectively. There are no the collapse of Einstein-Rosen bridge and Josephson oscillations of the scalar field.}
 \label{Fig:noCauchy}
\end{figure}
We will consider two particular cases. For the first example, we take $\beta=8$ and $A_{t}'(z_{H})=2$, for which the black hole~\eqref{BIbh} has a timelike singularity with a Cauchy horizon at $z_I\approx2.812z_H$. For the second one, we choose $\beta=2$ and $A_{t}'(z_{H})=1$, for which the solution~\eqref{BIbh} resembles a Schwarzschild like black hole without any inner horizon. Next, we turn on a small charged scalar with $\psi(z_H)=0.1$ and integrate the system from $z_H$ towards the singularity. In Fig.~\ref{Fig:Cauchy} we show the interior solutions for the first case. It is clear that at a value of $z$ close to the inner horizon of Born-Infeld black hole~\eqref{BIbh} $g_{tt}$ (green) shrinks rapidly and the scalar field $\psi$ (blue) goes through a series of oscillations. The later case is presented in Fig.~\ref{Fig:noCauchy}. As the normal Born-Infeld black hole now has no inner horizon, there is no similar instability triggered by the scalar hair as the first case of Fig.~\ref{Fig:Cauchy}. It is obvious that the collapse of Einstein-Rosen bridge and Josephson oscillations of the scalar hair vanish. Thus, we conclude that the intricate dynamics is associated with the instability of the would-be inner Cauchy horizon triggered by scalar hairs.

\section{Discussion}\label{Sec:discussion}
In this paper we have proved a no Cauchy horizon theorem for nonlinear electrodynamcs black holes with charged scalar hairs. We first extended the construction of the radially conserved quantity $\mathcal{Q}$ to the case with general NED. The resulted expression of $\mathcal{Q}$ depends explicitly on the form of NED, see~\eqref{Q}. Following~\cite{Cai:2020wrp}, we have shown that there exists no inner Cauchy horizon for both spherical and planar black holes with charged scalar hairs. For the hyperbolic horizon case, the method of~\cite{Cai:2020wrp} does not apply to the general NED. But we can show that for the hyperbolic case, the hairy black hole only has at most one inner horizon provided $L_s\geq0$. This result is based on the existence of the conserved charge $\mathcal{Q}$~\eqref{Q} and we have not imposed any constraint on the configuration of matter fields. Interestingly, by requiring the matter contents inside a black hole to satisfy the NEC, we found an easy way to rule out the existence of Cauchy horizon for all black holes with charged scalar hairs. This result is independent of the form of scalar potential $V$ and the electromagnetic sector $L(s)$ as well as the topology of a black hole. In particular, without any constraint on $L(s)$, there is no Cauchy horizon for the hyperbolic black hole case after considering the NEC. 

We have studied the interior dynamics in more details for the hairy Born-Infeld black hole in Section~\ref{Sec:BI}. Thanks to the special form of the Born-Infeld action~\eqref{BI}, we have shown that the geometry near the spacelike singularity takes a universal Kasner form when the kinetic term of the scalar dominates. We found that the scaling for $\chi$, $\psi$ and $f$ of the Born-Infeld theory takes the same form as the Maxwell case~\cite{Cai:2020wrp}. The difference happens for the gauge potential which has a different scaling due to the nonlinear structure of the electrodynamics. We then considered the effect of the Born-Infeld parameter $\beta$ on the black hole interior beyond the event horizon. For the normal Born-Infeld black hole~\eqref{BIbh}, its interior can be RN like with an inner Cauchy horizon or Schwarzschild like without Cauchy horizon, depending on $\beta$. We compared the interior dynamics for both cases and found that the intricate dynamics, including the collapse of Einstein-Rosen bridge and the Josephson oscillations, is closely related to the Cauchy horizon instability.

We conclude with a few comments given as follows. Our no Cauchy horizon result is robust in the sense that it is independent of the details of the action and the spacetime structure at the short distance. While we have limited ourselves to a minimal model~\eqref{action}, the theorem can be extended to cases with more nonminimal couplings. For example, we can modify the action to be 
\begin{equation}
\mathcal L_{M}=Z\,L(s)-K\,(D_{\mu}\Psi)^{*}(D^{\mu}\Psi)-V\,,
\end{equation}
with $Z$ and $K$ general functions of the scalar $\Psi$. The corresponding conserved charge is given by
\begin{equation}
\begin{aligned}
\mathcal{Q}(z)=&z^{2-d}e^{\chi/2}[z^{-2}(fe^{-\chi})'-Z\,L_{s}A_{t}A_{t}']\\
	     &+2k(d-1)\int^z y^{-d}e^{-\chi(y)/2}dy\,,
\end{aligned}
\end{equation}
Then it is easy to obtain the no Cauchy horizon theorem for the hairy black holes with spherical and planar topological horizons. After imposing the NEC, we can show that the combination $\frac{e^{\chi/2}}{z^{d-2}}Z\,L_{s} A_{t}' A_{t}$ is monotonic increasing with respect to $z$ and thus removes the inner Cauchy horizon. 

Although we have assumed $h$ to approach a negative constant at large $z$, \emph{i.e.} spacelike singularity, see~\eqref{hh}, the main discussion in Subsection~\ref{sub:kasner} in fact does not depend on whether the singularity is spacelike or not. A timelike singularity is possible for a hyperbolic black hole violating the NEC by fine tuning parameters. Following similar discussion, near the timelike singularity we can also obtain an asymptotic behavior which takes the same form of~\eqref{kasner}. The only difference is that $f_s$ now is a negative constant.  By performing the coordinate transformation $u \sim z^{-(1+\alpha^{2})/2}$, we obtain the following Lifshitz and hyperscaling violating solution:
\begin{equation}\label{lifsh}
\begin{split}
&ds^{2}=u^{\theta}\left(-\frac{dt^2}{u^{2\zeta}}+\frac{L^2 d u^2}{u^2}+\frac{d\Sigma^{2}_{2,k=-1}}{u^2}\right)\,,\\
&\psi(u)=-\frac{\sqrt{2}\alpha}{1+\alpha^2}\ln(u)\,,
\end{split}
\end{equation}
with $L$ a constant. Here the Lifshitz exponent is given by $\zeta=\frac{4}{1+\alpha^2}$ and the hyperscaling violating exponent $\theta=\frac{2(3+\alpha^2)}{1+\alpha^2}$. Given that the Born-Infeld action is well motivated from string theory, it deserves to understand possible physical consequence of the scaling geometry~\eqref{lifsh}.

While the argument with NEC is elegant in our present theory, the application of the NEC method is limited compared to the one based on the conserved charge $\mathcal{Q}$. Firstly, as we have mentioned, the conserved charge method does not impose any constraint on the energy momentum tensor $T_{\mu\nu}$ for matter fields. A well known example which essentially violates the NEC is  the traversable wormhole. Another one is the phantom energy which is a hypothetical form of dark energy explaining the accelerated expansion of our Universe and even seems to be favoured by observations~\cite{Aghanim:2018eyx,Jimenez:2016sgs,Vagnozzi:2018jhn,Alam:2016wpf}. Secondly, the success of the NEC method needs to relate the right hand side of~\eqref{nullAt} to the NEC. While it works well for the scalar with the present form, the generalization to new interactions or matter fields is not guaranteed. Lastly, there are many modified gravity theories which contain nonminimal couplings that directly couple curvature terms to matter fields, \emph{e.g.} $|\Psi|^2\mathcal{R}_{\mu\nu}\mathcal{R}^{\mu\nu}$ and $|\Psi|^2\mathcal{R}_{\mu\nu\gamma\delta}\mathcal{R}^{\mu\nu\gamma\delta}$. For these theories, there is no clear definition of $T_{\mu\nu}$ for which the NEC will be imposed~\cite{footnote}. In contrast, it has been recently shown that for general static spacetimes, one can always obtain the radially conserved quantity using the geometrical construction~\cite{Yang:2021civ}. Then one could continue the no Cauchy horizon argument following~\cite{Cai:2020wrp}.

The interior dynamics of the Born-Infeld theory was discussed for a simple quadratic potential which yields a Kasner form near the spacelike singularity. As the collapse and oscillations are nonlinear regimes that are likely sensitive to the potential, it is interesting to consider more general scalar potentials. In particular, it is worth uncovering some behaviors different from the Kasner form. In the present study of the interior dynamics, we have limited ourselves to the Born-Infeld black hole. It would be interesting to investigate other forms of NED, in particular, the ones that allow a non-singular black hole where the curvature invariants are regular everywhere. So far, we have considered the static black holes with a symmetric horizon. It would be a challenging question whether the theorem can be generalized to the stationary or inhomogeneous solutions.

\section*{Acknowledgement}
We would like to thank Rong-Gen Cai for helpful discussions and comments on the manuscript. This work was supported in part by the National Key Research and Development Program of China Grant No.2020YFC2201501, by the National Natural Science Foundation of China Grants No.12075298, No.11991052 and No.12047503 and by the Key Research Program of the Chinese Academy of Sciences Grant NO. XDPB15.

\appendix

\section{Hyperbolic black hole with \texorpdfstring{$L_s\geqslant0$}{TEXT}\label{App:Ls}}
\renewcommand{\theequation}{A\arabic{equation}}
In this section we prove that for the hyperbolic black hole with charged scalar hairs, there is at most one inner horizon behind the event horizon when $L_s\geqslant0$. Following the method of~\cite{Cai:2020wrp}, we first note that $z_{I}$ is the single root of $f(z)$. Otherwise one has $f''(z_I)\leqslant 0$ with $f'(z_I)=0$. Then we find 
\begin{equation}
\begin{aligned}
\mathcal Q'(z_{I})=\frac{f''(z_{I})}{z_{I}^{d}}&e^{-\chi(z_{I})/2}-2(d-1)z_{I}^{-d} e^{-\chi/2}\\
-&L_{s}z_{I}^{2-d}e^{\chi(z_{I}/2)}A_{t}'(z_{I})^2\,.
\end{aligned}
\end{equation}
It is clear that $\mathcal Q'(z_{I})<0$ as $L_s\geqslant 0$. This is obviously contradictory to our previous result $\mathcal Q'(z)=0$. Therefore, $z_{I}$ should be a single root with $f'(z_{I})>0$.

Secondly, assume that there is a second inner horizon arising at $z=z_{J}> z_{I}$. We will have $f(z_{J})=0$ and $f'(z_{J})\leq0$. With two inner horizons at $z_{I}$ and $z_{J}$, we have
\begin{equation}\label{qij}
\begin{aligned}
\frac{f'(z_{I})}{z_{I}^{d}}&e^{-\chi(z_{I})/2}-\frac{f'(z_{J})}{z_{J}^{d}}e^{-\chi(z_{J})/2}\\
\qquad&=-2(d-1) \int_{z_{I}}^{z_{J}} y^{-2}e^{-\chi(y)/2} dy.
\end{aligned}
\end{equation}
for the hyperbolic case $k=-1$. One immediately finds that both sides of~\eqref{qij} have different sign. Therefore, the second inner horizon will not form.

\section{No divergence of \texorpdfstring{$h(z)$}{TEXT} near the singularity}\label{App:h}
\renewcommand{\theequation}{B\arabic{equation}}
In this section we will show that for the Born-Infeld theory $h(z)$ near the singularity $z\rightarrow\infty$ is not divergent when the contribution of scalar potential $V$ can be neglected. We point out that the discussion below will not depend on whether the singularity is spaclike or not.

We begin with assuming that $h$ is not bounded when $z$ goes to infinity. Thanks to the structure of Born-Infeld action~\eqref{BI} which takes
\begin{equation}\label{app:BI}
L(s)=\beta^{2}\left(1-\sqrt{1-\frac{z^4 e^\chi A_t'^2}{\beta^{2}}}\right),
\end{equation}
with the ansatz~\eqref{ansatz}. One immediately obtains that
\begin{equation}\label{app:At}
A_{t}'^{2} < \frac{e^{-\chi} \beta^{2}}{z^{4}}\,.
\end{equation}
From~\eqref{eqchi} one finds that $\chi$ is a monotonic increasing function and is finite at $z_H$. Therefore, we have $e^{-\chi/2}\leqslant \mathcal{O}(z^0)$. 
Then we can obtain from~\eqref{app:At} that $|A_{t}|$ is at most of $\mathcal{O}(z^0)$. We point out that in contrast to the Maxwell theory, for the Born-Infeld case we are able to give a strong constraint on the magnitude of the gauge potential by simply requiring the Born-Infeld action~\eqref{app:BI} to be real.
So the last term of~\eqref{eqpsi} can be neglected and one has 
\begin{equation}\label{app:psi}
\psi''=-\left(\frac{h'}{h}+\frac{1}{z}\right)\psi'\,.
\end{equation}

Integrating~\eqref{app:psi}, one obtains
\begin{equation}
\psi(z)=\psi(z_0)+\psi'(z_0) z_0 h(z_0)\int_{z_0}^z  \frac{1}{x\, h(x)} dx\,,
\end{equation}
with $z_0$ a radial position sufficiently closed to the singularity.
Note that
\begin{equation}\label{app:opsi}
\Big|\int_{z_0}^z  \frac{1}{x\, h(x)} dx\Big| \leqslant  \int_{z_0}^z  \frac{1}{|x\, h(x)|} dx<\int_{z_0}^z  \frac{1}{x} dx=\ln(z-z_0)\,,
\end{equation}
where we have used our initial assumption that $h$ is sufficiently large at large $z$. Therefore, we find that $\mathcal{O}(|\psi|) <\mathcal{O}(\ln(z))$ near the singularity.

We can estimate from~\eqref{eqrho} that
\begin{equation}
|\rho'|=|\frac{2q^{2}\psi^{2}A_{t}}{z^{5}h}|<|\frac{q^{2}\ln(z)^2}{z^{5}}|\rightarrow 0\,,
\end{equation}
where have have used $\mathcal{O}(|\psi|) <\mathcal{O}(\ln(z))$ and $|A_{t}|$ is at most of $\mathcal{O}(z^0)$ at large $z$.
Then, we have
\begin{equation}\label{app:rho}
\rho=L_{s} e^{\chi/2}A_{t}'=\frac{e^{\chi/2}A_t'}{\sqrt{1-\frac{z^4 e^\chi A_t'^2}{\beta^2}}}\approx \rho_0,
\end{equation}
with $\rho_0$ a constant. It is easy to get from~\eqref{app:rho} that 
\begin{equation}
e^{\chi}A_t'^2=\frac{\beta^2}{z^4+\beta^2/\rho_0^2}\,,
\end{equation}
and therefore
\begin{equation}
L_{s} =\sqrt{\frac{\rho_0^2}{\beta^2}z^4+1}\sim \mathcal{O} (z^2)\,,
\end{equation}
at large $z$. Hence, we finally obtain that
\begin{equation}
h'\sim\frac{e^{-\chi/2}}{2}\left(-\frac{2k}{z^{2}}-\frac{\beta^{2}}{z^{4}}+\frac{\beta^{2}}{z^{4}}L_{s}\right)\sim \mathcal{O}(z^{-2}).
\end{equation}
It's obvious that $h'$ is integrable which is not consistent with our assumption that $h$ is divergent near the singularity. Thus, we conclude that $h$ is bounded as approaching to the singularity.

\section{Some examples of scalar potential}\label{App:V}
In Subsection~\ref{sub:kasner} we have obtained the Kasner form near the spacelike singularity with the assumption that the scalar potential can be neglected at large $z$. In this section we will consider two typical kinds of scalar potential without above assumption a priori, including a polynomial potential and an exponential form. For the polynomial potential we can explicitly show that its contribution at large $z$ can be neglected. In contrast, one has to consider the contribution from the exponential potential at large $z$.

\renewcommand{\theequation}{C\arabic{equation}}

\subsection{Polynomial potential}

For the scalar potential in terms of finite polynomials of $|\Psi|^2$, we assume that its dominant term is $V_n=|\Psi|^{2n}$ with $n$ a constant. We will show that the contribution from $V_n$ to the equations of motion~\eqref{eqh} to~\eqref{eqrho} is subdominated at large $z$.
can be neglected at large $z$. There are two cases for $\psi^2$. The first one is $\mathcal{O}(\psi^2) \leqslant \mathcal{O}(\ln(z))$ for which it is easy to show that its contribution is subdominated. Therefore, we consider the second case for which $\mathcal{O}(\psi^2) > \mathcal{O}(\ln(z))$. We will prove that $V_n e^{-\chi/2}$ decreases monotonically at large $z$ and thus is bounded.

When $\mathcal{O}(\psi^2) > \mathcal{O}(\ln(z))$, one can find that
\begin{equation}
    \lim_{z\to \infty} (\frac{\psi^2A_t^2q^2}{h^2z^5})/(z\psi'^2) < \lim_{z\to \infty} \frac{A_t^2 q^2}{h^2} \frac{(\ln{z})^2}{z^4}=0\,,
\end{equation}
where we have used that at large $z$ $|A_{t}|$ is at most of $\mathcal{O}(z^0)$ (see~\eqref{app:At}), $\mathcal{O}((\ln|\psi|)') > \mathcal{O}(\frac{1}{z\ln(z)})$ and~\eqref{hh}. 
Therefore, the last term of~\eqref{eqchi} can be neglected comparing to the $z\psi'^2$ term. 
Then, we can derive
\begin{equation}\label{appV}
    (V_n e^{-\chi/2})'=-\frac{(\psi^2)'}{2\psi^2}(z(\psi^2)'-2n)(V_n e^{-\chi/2})<0\,,
\end{equation}
for $\mathcal{O}(\psi^2) > \mathcal{O}(\ln(z))$ at large $z$. Therefore $V_n e^{-\chi/2}$ is bounded. That means the potential term in~\eqref{eqh} and~\eqref{eqpsi} can be neglected near the singularity. Indeed, one can explicitly check that for the Kanser form~\eqref{kasner} with $\psi\sim\ln(z)$,~\eqref{appV} is satisfied for general polynomial potentials.

\subsection{Exponential potential}
For the exponential potential, we heuristically consider $V_e=e^{\kappa |\Psi|^{2}}$ with $\kappa$ a positive constant. We now show that one can not ignore the contribution from the scalar potential $V_e$ at large $z$.

Suppose the scalar potential can be neglected at large $z$. Our discussion in Subsection~\ref{sub:kasner} applies and one obtains the Kasner form~\eqref{kasner} with $\psi=\sqrt{2}\alpha\ln(z)$ and $\chi=2\alpha^2\ln(z)$ at large $z$. Then we have
\begin{equation}
 V_e e^{-\chi/2}=e^{2\kappa \alpha^2\ln(z)^2}z^{-\alpha^2}\,,
\end{equation}
which is much larger than the contribution from the kinetic term of $\psi$ as well as gauge potential. Thus, for the case with exponential potential, one should consider the potential term near the singularity. It is also obvious that $V_e$ violates the constraint~\eqref{kconstraint} at large $z$ and thus novel behaviors different from the Kasner form could appear.



\begin{thebibliography}{99}
 \small

\bibitem{Penrose:1964wq}
R.~Penrose, Gravitational collapse and space-time singularities,''
Phys. Rev. Lett. \textbf{14} (1965) 57--59.

\bibitem{Hawking:1969sw}
S.~W. Hawking and R.~Penrose, ``The Singularities of gravitational collapse and cosmology," Proc.Roy. Soc. Lond. A \textbf{314} (1970) 529--548.
  
\bibitem{GWs:2016}
B.~Abbott \emph{et al.}, ``Observation of Gravitational Waves from a Binary Black Hole Merger,"
Phys. Rev. Lett. \textbf{116}(6)(2016)061102.

\bibitem{shadow:2019I}
K.~Akiyama \emph{et al.}, ``First M87 Event Horizon Telescope Results. I. The Shadow of the Supermassive Black Hole,"
 Astrophys. J. \textbf{L1}(1)(2019)875.

 \bibitem{shadow:2019IV}
 K.~Akiyama \emph{et al.}, ``First M87 Event Horizon Telescope Results. IV. Imaging the Central Supermassive Black Hole,"
  Astrophys. J. \textbf{L4}(1)(2019)875.
  
\bibitem{Geroch:1969}
Y.~Choquet-Bruhat and R.~P.~Geroch, ``Global Aspects of the Cauchy Problem in General Relativity,'' Commun.~Math.~Phys.~14,~329(1969).
\bibitem{Ringstrom:2015jza}
H.~Ringstr\"{o}m,
``Origins and development of the Cauchy problem in general relativity,''
Class. Quant. Grav. \textbf{32}, no.12, 124003 (2015).

\bibitem{Isenberg:2015rqa}
J.~Isenberg,
``On Strong Cosmic Censorship,''
[arXiv:1505.06390 [gr-qc]].

\bibitem{Hartnoll:2020rwq}
S.~A.~Hartnoll, G.~T.~Horowitz, J.~Kruthoff and J.~E.~Santos,
``Gravitational duals to the grand canonical ensemble abhor Cauchy horizons,''
JHEP \textbf{10}, 102 (2020)
[arXiv:2006.10056 [hep-th]].

\bibitem{Hartnoll:2020fhc}
S.~A.~Hartnoll, G.~T.~Horowitz, J.~Kruthoff and J.~E.~Santos,
``Diving into a holographic superconductor,''
SciPost Phys. \textbf{10}, 009 (2021)
[arXiv:2008.12786 [hep-th]].

\bibitem{Cai:2020wrp}
R.~G.~Cai, L.~Li and R.~Q.~Yang,
``No Inner-Horizon Theorem for Black Holes with Charged Scalar Hairs,''
JHEP \textbf{03}, 263 (2021)
[arXiv:2009.05520 [gr-qc]].

\bibitem{Devecioglu:2021xug}
D.~O.~Devecioglu and M.~I.~Park,
``No Scalar-Haired Cauchy Horizon Theorem in Einstein-Maxwell-Horndeski Theories,''
[arXiv:2101.10116 [hep-th]].

\bibitem{Grandi:2021ajl}
N.~Grandi and I.~Salazar Landea,
``Diving inside a hairy black hole,''
[arXiv:2102.02707 [gr-qc]].




\bibitem{Born:1934gh}
M.~Born and L.~Infeld, ``Foundations of the new field theory,'' Proc. Roy. Soc. Lond. A \textbf{144} no. 852, (1934) 425--451.

\bibitem{Heisenberg:1935qt}
W.~Heisenberg and H.~Euler,
``Consequences of Dirac's theory of positrons,''
Z. Phys. \textbf{98}, no.11-12, 714-732 (1936)
[arXiv:physics/0605038 [physics]].

\bibitem{Bandos:2020jsw}
I.~Bandos, K.~Lechner, D.~Sorokin and P.~K.~Townsend,
``A non-linear duality-invariant conformal extension of Maxwell's equations,''
Phys. Rev. D \textbf{102}, 121703 (2020)
[arXiv:2007.09092 [hep-th]].

\bibitem{AyonBeato:1998ub}
E.~Ayon-Beato and A.~Garcia,
``Regular black hole in general relativity coupled to nonlinear electrodynamics,''
Phys. Rev. Lett. \textbf{80}, 5056-5059 (1998)
[arXiv:gr-qc/9911046 [gr-qc]].

\bibitem{Kiritsis:2016cpm}
E.~Kiritsis and L.~Li,
``Quantum Criticality and DBI Magneto-resistance,''
J. Phys. A \textbf{50}, no.11, 115402 (2017)
[arXiv:1608.02598 [cond-mat.str-el]].

\bibitem{Blauvelt:2017koq}
E.~Blauvelt, S.~Cremonini, A.~Hoover, L.~Li and S.~Waskie,
``Holographic model for the anomalous scalings of the cuprates,''
Phys. Rev. D \textbf{97}, no.6, 061901 (2018)
[arXiv:1710.01326 [hep-th]].

\bibitem{Cremonini:2018kla}
S.~Cremonini, A.~Hoover, L.~Li and S.~Waskie,
``Anomalous scalings of cuprate strange metals from nonlinear electrodynamics,''
Phys. Rev. D \textbf{99}, no.6, 061901 (2019)
[arXiv:1812.01040 [hep-th]].



\bibitem{Camara:2004ap}
C.~S.~Camara, M.~R.~de Garcia Maia, J.~C.~Carvalho and J.~A.~S.~Lima,
``Nonsingular FRW cosmology and nonlinear electrodynamics,''
Phys. Rev. D \textbf{69}, 123504 (2004)
[arXiv:astro-ph/0402311 [astro-ph]].

\bibitem{Kruglov:2015fbl}
S.~I.~Kruglov,
``Universe acceleration and nonlinear electrodynamics,''
Phys. Rev. D \textbf{92}, no.12, 123523 (2015)
[arXiv:1601.06309 [gr-qc]].

\bibitem{Ovgun:2017iwg}
A.~\"Ovg\"un, G.~Leon, J.~Maga\~na and K.~Jusufi,
``Falsifying cosmological models based on a non-linear electrodynamics,''
Eur. Phys. J. C \textbf{78}, no.6, 462 (2018)
[arXiv:1709.09794 [gr-qc]].

\bibitem{Benaoum:2021tec}
H.~B.~Benaoum and A.~Ovgun,
``Matter-antimatter asymmetry induced by non-linear electrodynamics,''
[arXiv:2105.07695 [gr-qc]].


\bibitem{Bokulic:2021dtz}
A.~Bokuli\'c, T.~Juri\'c and I.~Smoli\'c,
``Black hole thermodynamics in the presence of nonlinear electromagnetic fields,''
[arXiv:2102.06213 [gr-qc]].

\bibitem{Nojiri:2017kex}
S.~Nojiri and S.~D.~Odintsov,
``Regular multihorizon black holes in modified gravity with nonlinear electrodynamics,''
Phys. Rev. D \textbf{96}, no.10, 104008 (2017)
[arXiv:1708.05226 [hep-th]].

\bibitem{Gao:2017vqv}
C.~Gao, Y.~Lu, S.~Yu and Y.~G.~Shen,
``Black hole and cosmos with multiple horizons and multiple singularities in vector-tensor theories,''
Phys. Rev. D \textbf{97}, no.10, 104013 (2018)
[arXiv:1711.00996 [gr-qc]].

\bibitem{Hendi:2015esa}
S.~H.~Hendi and M.~Momennia,
``Thermodynamic instability of topological black holes with nonlinear source,''
Eur. Phys. J. C \textbf{75}, no.2, 54 (2015)
[arXiv:1501.04863 [gr-qc]].

\bibitem{footnote0}
It is possible that $z_I$ is a local maximum of $f(z)$ for which $f'(z_I)=0$. This will not change our proof of no Cauchy horizon.



\bibitem{Aghanim:2018eyx}
N.~Aghanim \textit{et al.} [Planck],
``Planck 2018 results. VI. Cosmological parameters,''
Astron. Astrophys. \textbf{641}, A6 (2020)
[arXiv:1807.06209 [astro-ph.CO]].

\bibitem{Jimenez:2016sgs}
J.~Beltran Jimenez, R.~Lazkoz, D.~Saez-Gomez and V.~Salzano,
``Observational constraints on cosmological future singularities,''
Eur. Phys. J. C \textbf{76}, no.11, 631 (2016)
[arXiv:1602.06211 [gr-qc]].

\bibitem{Vagnozzi:2018jhn}
S.~Vagnozzi, S.~Dhawan, M.~Gerbino, K.~Freese, A.~Goobar and O.~Mena,
``Constraints on the sum of the neutrino masses in dynamical dark energy models with $w(z) \geq -1$ are tighter than those obtained in $\Lambda$CDM,''
Phys. Rev. D \textbf{98}, no.8, 083501 (2018)
[arXiv:1801.08553 [astro-ph.CO]].

\bibitem{Alam:2016wpf}
U.~Alam, S.~Bag and V.~Sahni,
``Constraining the Cosmology of the Phantom Brane using Distance Measures,''
Phys. Rev. D \textbf{95}, no.2, 023524 (2017)
[arXiv:1605.04707 [astro-ph.CO]].

\bibitem{Yang:2021civ}
R.~Q.~Yang, R.~G.~Cai and L.~Li,
``Constraining the number of horizons with energy conditions,''
[arXiv:2104.03012 [gr-qc]].

\bibitem{footnote}
One could introduce the ``effective" energy momentum tensor assuming the standard Einstein's equation.  More precisely, one groups all the dependence of the Einstein tensor $G_{\mu\nu}$ on the left hand side of Einstein's equation. Nevertheless, the violation of the NEC from such ``effective" energy momentum tensor indeed can be possible, see \emph{e.g.}~\cite{Mandal:2019}.

\bibitem{Mandal:2019}
S.~Mandal,
``Revisiting Laws of Black Hole Mechanics and Violation of Null Energy Condition,"
Journal of High Energy Physics, Gravitation and Cosmology, 5, 82-111 (2019).


\end{thebibliography}
\end{document}